\documentclass[%
 aip,
 jmp,%
 amsmath,amssymb,
 reprint,%
]{revtex4-1}
{\usepackage{amsmath}
\usepackage{braket}
\usepackage{color}
\usepackage{bm}
\usepackage{graphicx}

\renewcommand{\b}[1]{\bar{{#1}}}

\begin{document}

\title{A transformed framework for dynamic correlation in multireference problems}

\begin{abstract}
We describe how multirefence dynamic correlation theories 
can be naturally obtained as single-reference correlation
theories in a canonically transformed frame. Such canonically transformed  correlation theories
are very simple and involve identical expressions to their
single-reference counterparts. The corresponding
excitations involve quasiparticles rather than the bare
particles of the system. 
High-order density matrices
(or their approximations) and the numerical metric instabilities 
common to multireference correlation theories do not appear. As an example,
we formulate the Bogoliubov canonically transformed version of second-order M\o ller-Plesset perturbation theory
and demonstrate its performance in H$_2$, H$_2$O, N$_2$, and BeH$_2$ bond dissociation.
\end{abstract}
\author{Alexander Yu.\@ Sokolov}
\email{alexsokolov@princeton.edu}
\author{Garnet Kin-Lic Chan}
\email{gkchan@princeton.edu}
\affiliation{Department of Chemistry, Princeton University, Princeton, NJ 08544}
\maketitle

\section{Introduction}
Multireference correlation remains a driving
force for the development of new quantum chemical methods. Typically
the orbitals are divided into two sets: an active space
with near-degeneracies and an external set
of empty or core orbitals. It is now possible to describe the
 correlation in the active orbitals for active spaces
with up to 50 orbitals, to produce a multireference active space wavefunction $\ket{\Psi_0}$
that is formally the sum of many determinants.\cite{Olsen:1988p2185,Malmqvist:1990p5477,Piecuch:1999p6103,White:1999p4127,Olsen:2000p7140,Kallay:2001p2945,Legeza2008,Booth:2009p054106,Kurashige:2009p234114,Marti:2011p6750,Small:2011p19285,Chan:2011p465,Booth:2012p164112,Kurashige:2013p660,Wouters:2014p272,Booth:2014p1855,Sharma:2014p927,Evangelista:2014p124114}
In these cases, the remaining challenge is  to efficiently describe the
correlation outside of the active space, involving the external orbitals.
We refer to this as the dynamic correlation problem in a multireference setting.

Dynamic correlation from a {\it single} reference (single determinant) can
be considered well understood, and is well-captured by low-order
perturbation theory (such as M\o ller-Plesset perturbation theory),\cite{Szabo:1982,Helgaker:2000} configuration interaction,\cite{Szabo:1982,Helgaker:2000}
or coupled cluster methods.\cite{Crawford:2000p33,Bartlett:2007p291,Shavitt:2009} 
Analogues of these methods for multireference problems, such as multireference perturbation theory,\cite{Wolinski:1987p225,Andersson:1992p1218,Hirao:1992p374,Werner:1996p645,Finley:1998p299,Kurashige:2011p094104}
multireference configuration interaction,\cite{Buenker:1974p33,Siegbahn:1980p1647,Werner:1988p5803,Saitow:2013p044118} and multireference coupled cluster and canonical transformation\cite{Mukherjee:1977p955,Lindgren:1978p33,Jeziorski:1981p1668,Mahapatra:1998p6171,Yanai:2006p194106,Evangelista:2007p024102,Yanai:2007p104107,Datta:2011p214116,Evangelista:2011p114102,Kohn:2012p176,Nooijen:2014p081102} methods
have also been formulated. However, all of these
multireference formulations are algebraically more opaque and computationally much more
expensive than their single-reference counterparts.

Here we define a natural framework to construct multireference
dynamic correlation methods with precisely the same equations, and same complexity,
as existing well-known and well-understood single-reference theories. The idea
is to consider dynamic correlation within the frame of {\it canonically
transformed} interactions and quasiparticles. Within such a view, the 
multireference initial state is viewed as a ``vacuum'' 
of the quasiparticles, and the Hamiltonian is expressed in terms of a modified
set of ``integrals''. Once these integrals are defined, the correlation
treatment is {\it precisely} that of a single-reference theory. Thus complications
common to multireference methods, such as high-order
density matrices and singular metrics, do not appear. Similar motivations
have led to Mukherjee and Kutzelnigg's earlier formulation of the generalized normal
ordering and Wick's theorem.\cite{Mukherjee:1997p561, Kutzelnigg:1997p432} However, as we shall see, our framework is different, leading to
different formalism or methods. As this paper was finalized for submission, Rolik and K\'{a}llay  published work\cite{Rolik:2014p134112} with a similar conceptual foundation to our own,
although differing in technical details. The relationship between the two
is discussed below.

\section{Recap of canonical transformations}

We will use the concept of canonical transformations. To 
improve understanding, we recall some salient points here, and a complete
discussion may be found in standard texts.\cite{Ring1980,Blaizot1986} We
first work with a concrete basis of creation and annihilation operators, $c$ and $c^\dag$. 
Any two normalized states are related by a many-particle canonical (i.e. unitary) transformation $\hat{U}$ with $\hat{U}\hat{U}^\dag=1$. For example, 
a multireference state, a sum of many determinants, can be related to a single 
determinant,
\begin{align}
|\Psi\rangle = \hat{U} |\mathrm{det}_c\rangle \ . \label{eq:ctdet}
\end{align}
In Eq.~(\ref{eq:ctdet}), the operator $\hat{U}$ is particle-number-conserving. The transformation may formally 
be parametrized in exponential form, $\hat{U}=\exp \hat{A}$, where $\hat{A}$
is expanded as
\begin{align}
\hat{A}=\sum_{pq} A_{pq} c^\dag_p c_q + \sum_{pqrs} A_{pqrs} c^\dag_p c^\dag_q c_r c_s + \ldots
\label{eq:n_conserv}
\end{align}
with elements of antihermitian tensors $A_{pq}$, $A_{pqrs}$, etc.  

A canonical transformation need not be particle-number-conserving. For example, a multireference state can also be
related to a vacuum,
\begin{align}
|\Psi\rangle = \hat{U} |\mathrm{vac}_c\rangle \ . \label{eq:ctvac}
\end{align}
In this case, the operator $\hat{A}$ can be written as
\begin{align}
\label{eq:n_nonconserv}
\hat{A} &= \sum_{pq} A_{pq} c^\dag_p c_q + B_{pq} (c^\dag_p c^\dag_q + c_q c_p) \notag\\&
+  \sum_{pqrs} A_{pqrs} c^\dag_p c^\dag_q c_r c_s + B_{pqrs} (c^\dag_p c^\dag_q c^\dag_r c_s + c^\dag_s c_r c_q c_p) \notag\\&
+ \sum_{pqrs} C_{pqrs} (c^\dag_p c^\dag_q c^\dag_r c^\dag_s + c_s c_r c_q c_p) + \ldots
\end{align}
where tensors $B_{pq}$, $B_{pqrs}$, etc.\@ are also antihermitian. 
The state $\ket{\mathrm{vac}_c}$ in Eq. \eqref{eq:ctvac} is the vacuum of the $c$ operators satisfying the relationship
\begin{align}
c\ket{\mathrm{vac}_c}=0
\end{align}
for any $c$.
The particle-number-conserving transformation \eqref{eq:ctdet} and \eqref{eq:n_conserv} is thus a special case of a more general non-number-conserving transformation in Eqs. \eqref{eq:ctvac} and \eqref{eq:n_nonconserv}. We can therefore denote both transformations as $\hat{U}$, without a loss of generality. Note though that
an exponential parametrization is {\it not 
essential} to the definition of $\hat{U}$. In fact, we do not use such a parametrization in our numerical work below.

Having introduced the concept of a canonical transformation for a multireference state $\ket{\Psi}$, we now discuss canonical transformations with respect to the individual creation and annihilation operators (e.g., $c$ and $c^\dag$). Let us first consider the case of the number-conserving canonical transformation $\hat{U}$ in Eq.~(\ref{eq:ctdet}). We recall that a single determinant can be expressed as a polynomial of creation operators acting on an appropriate vacuum, i.e. $\ket{\mathrm{det_c}} = c_1^\dag c_2^\dag \ldots  c^\dag_N \ket{\mathrm{vac}_c}$. The multireference state $\ket{\Psi}$ can now be written in the following form:
 \begin{align}
\label{eq:qp_transformation}
 \ket{\Psi(c,c^\dag)} &= 
 \hat{U} \ket{\mathrm{det}_c}\notag\\
&= \hat{U} c_1^\dag \hat{U}^\dag \hat{U}  c_2^\dag  \ldots  c^\dag_N \hat{U}^\dag \hat{U}\ket{\mathrm{vac}_c} \notag\\
&= a_1^\dag a_2^\dag \ldots  a_N^\dag \hat{U} \ket{\mathrm{vac}_c} \notag\\
&= a_1^\dag a_2^\dag \ldots  a_N^\dag  \ket{\mathrm{vac}_a} \notag\\
&\equiv \ket{\mathrm{det}_a} = \ket{\Psi(a,a^\dag)} \ .
\end{align}
Eq. \eqref{eq:qp_transformation} demonstrates that a multideterminant $N$-particle state $\ket{\Psi}$ expressed in the frame of $c$ and $c^\dag$ operators (i.e. as a polynomial of $c$ and $c^\dag$) can be written as a single determinant in a new $a, a^\dag$ frame, involving a simple product of $a^\dag$ operators.
Here the notation $\ket{\Psi(c,c^\dag)}$, $\ket{\Psi(a,a^\dag)}$ is
used to denote that these are the {\it same} states, only expressed
as different functions of the underlying $c, c^\dag$, and $a, a^\dag$ bases.
The notation $\ket{\mathrm{vac}_c}$, $\ket{\mathrm{vac}_a}$ however 
denotes that the vacua of $c, c^\dag$ and $a, a^\dag$ operators are {\it different}, as they are related by $\hat{U}$.

In Eq. \eqref{eq:qp_transformation}, the canonically transformed quasiparticle operators $a,a^\dag$ are defined as~\cite{Blaizot1986,Neuscamman:2010p231}
\begin{align}
a^{(\dag)}=\hat{U}c^{(\dag)} \hat{U}^\dag \ ,
\label{eq:a_def}
\end{align}
where $a^{(\dag)}$ represents either $a^{\dag}$ or $a$. Since $\hat{U}\hat{U}^\dag=1$,
the operators $a^{(\dag)}$ then have the same commutation properties as those of $c^{(\dag)}$, e.g.
\begin{align}
a^{}_p a^\dag_q + a^\dag_q a^{}_p = \delta_{pq} \ . 
\label{eq:a_commut}
\end{align}
Analogously, for a more general canonical transformation in Eq.~(\ref{eq:ctvac}), we can 
regard the general state in the $c,c^\dag$ frame as the {\it vacuum} in the $a,a^\dag$ frame:
\begin{align}
\ket{\Psi(c,c^\dag)} &= \hat{U} \ket{\mathrm{vac}_c} = \ket{\mathrm{vac}_a} \ , \notag\\
a\ket{\mathrm{vac}_a}&=0 \ .
\label{eq:ctqpvac}
\end{align}

The simplest canonical transformation is a single-particle transformation. The special particle-number-conserving case is an orbital rotation, 
corresponding to $\hat{A}=\sum_{pq} A_{pq} c^\dag_p c_q$, where the amplitudes $A_{pq}$
are the elements of a matrix $\mathbf{A}$. The quasiparticle operators $a^{(\dag)}_p$ are then expressed as a linear transformation
\begin{align}
a^{(\dag)}_p = \sum_{q} \alpha_{pq} c^{(\dag)}_q \ , \label{eq:sp}
\end{align}
where the matrix $\bm{\alpha}=\exp\mathbf{A}$ and $\bm{\alpha}\bm{\alpha}^\dag=\bm{1}$. 
In spin-restricted form, $\bm{\alpha}$ is the same for up or down spin.
The general single-particle transformation (a Bogoliubov transformation)\cite{Bogoliubov:1947p23} is not number-conserving, corresponding to the first two terms in Eq.~(\ref{eq:n_nonconserv}). In this case, the quasiparticle operators are given by the  general linear transformation
\begin{align}
\label{eq:bogoliubov_general}
a_{p} &= \sum_{q}\alpha_{pq} c_{q} + \beta_{pq} c^\dagger_q\ , 
\end{align}
where the matrices satisfy $\bm{\alpha}^\dag \bm{\alpha} + \bm{\beta}^\dag \bm{\beta}=\bm{1}$  for unitarity. In the spin-restricted form, the Bogoliubov transformation becomes
\begin{align}
a_p = \sum_q \alpha_{pq} c_{q} + s_{p} \beta_{pq} c^\dag_{\bar{q}} \ ,
\label{eq:a_bogoliubov_restricted}
\end{align}
where $s_{p} = -1$ or $+1$ for the spin-orbital label $p$ with spin up or down, respectively; $p, q$ have the same spin; and $\bar{p}$  corresponds to $p$ with the opposite spin.  
Eq.~(\ref{eq:a_bogoliubov_restricted})  can be seen to be spin-restricted
because the total spin of the state changes in the same way either by creating
a particle of given spin ($c^\dag_{\bar{q}}$), 
or destroying a particle of opposite spin ($c_q$). The corresponding Bogoliubov vacuum $\ket{\mathrm{vac}_a}$ defined by Eq.~(\ref{eq:bogoliubov_general})
is equivalent to the famous Bardeen-Cooper-Schrieffer (BCS) state of superconductivity.\cite{Cooper:1956p1189,Bardeen:1957p162,Bardeen:1957p1175}
Finally, for an arbitrary canonical transformation, the quasiparticle operators are
polynomials in the bare particle operators $c^{(\dag)}$, thus including 
cubic, and higher terms,
\begin{align}
a_p &= \sum_{q} \alpha^{(1)}_{pq} c_q + \beta^{(1)}_{pq} c_q^\dag \notag\\&+ 
\sum_{qrs} \alpha^{(2)}_{pqrs} c_q^\dag c_r c_s + \beta^{(2)}_{pqrs} c_q^\dag c_r^\dag c_s +\ldots
\label{eq:ctqp}
\end{align}

Formally, it is entirely equivalent to work in the $c, c^\dag$ frame
or the  $a, a^\dag$ frame. To transform the computation
from the $c, c^\dag$ frame to the $a, a^\dag$ frame, we
must re-express operators (polynomials
of $c, c^\dag$) and states (polynomials of $c, c^\dag$ acting on $\ket{\mathrm{vac}_c}$),
as corresponding functions of $a,a^\dag$. 
Consider starting with the electronic
Hamiltonian in the bare basis $c, c^\dag$, 
\begin{align}
\hat{H}(c,c^\dag) = \sum_{pq} t_{pq} c^\dag_p c_q + \frac{1}{4} \sum_{pqrs} v_{pqrs} c^\dag_p c^\dag_q c_s c_r \ ,
\label{eq:h}
\end{align}
where $t_{pq}$ and $v_{pqrs}$ are the usual one- and antisymmetrized two-electron integrals, respectively. To work in the transformed frame, we rewrite this
in terms of $a, a^\dag$ using the inverse of Eq. \eqref{eq:ctqp}, yielding
\begin{align}
\label{eq:h_ct}
\hat{H}(a, a^\dag) 
&= \sum_{pq} t^{(\dag,\dag)}_{pq} a^{(\dag)}_p a^{(\dag)}_q + 
\sum_{pqrs} v^{(\dag,\dag,\dag,\dag)}_{pqrs} a^{(\dag)}_p a^{(\dag)}_q a^{(\dag)}_r a^{(\dag)}_s \notag\\
&+ \sum_{pqrstu} w_{pqrstu}^{(\dag,\dag,\dag,\dag,\dag,\dag)}
a^{(\dag)}_p a^{(\dag)}_q a^{(\dag)}_r a^{(\dag)}_s a^{(\dag)}_t a^{(\dag)}_u  +\ldots.
\end{align}
Higher-body terms and non-particle-number-conserving terms naturally appear in the case of a general transformation \eqref{eq:ctqp}, since the r.h.s. of Eq.~(\ref{eq:ctqp}) is both non-linear and non-number-conserving.
Note the notation $\hat{H}(a,a^\dag)$  indicates that it is the {\it same} Hamiltonian as $H(c,c^\dag)$, only expressed in terms of different particles.
As $a(c,c^\dag)$ and $c(a,a^\dag)$ are defined in terms of the $\alpha$ and $\beta$
coefficients in Eq.~(\ref{eq:ctqp}),
the integrals in Eqs.~(\ref{eq:h}) and (\ref{eq:h_ct}) are related by
these coefficients. For a number-conserving single-particle transformation,
this relationship is  the standard integral orbital transformation. The expressions for a Bogoliubov transformation are given in the Appendix.

\section{Multireference as single-reference in the transformed frame}

The basic premise of this work is as follows:  we are free to work either with the bare particles ($c, c^\dag$) or the quasiparticles ($a, a^\dag$), thus we can choose the most convenient representation. 
In the case of a multireference correlation theory, we typically have a multideterminant reference $\ket{\Psi_0}$ defined in a space of core (doubly occupied) and active orbitals. 
To describe dynamic correlation, excitations between these sets of orbitals and a set of external (unoccupied) orbitals need to be included. 
The multideterminantal structure of $\ket{\Psi_0}$ in the $c,c^\dag$ frame gives rise to complicated expressions for the matrix elements of operators, which include up to $n$-body reduced density matrices for an $n$-particle operator.\cite{Kutzelnigg:1997p432}
On the other hand, in the transformed $a,a^\dag$ frame, the reference state $\ket{\Psi_0}$ appears simpler, such as a determinant or a vacuum of quasiparticles (Eqs. \eqref{eq:qp_transformation} or \eqref{eq:ctqpvac}), and the corresponding matrix elements of operators have single-reference form. 
Working in the transformed frame requires a more complicated form of the Hamiltonian $\hat{H}(a,a^\dag)$ (Eq.~\ref{eq:h_ct}). However, once the quasiparticle transformation is performed, all equations for the multireferenence dynamic correlation theory in the transformed $a, a^\dag$ frame are identical to the single-reference theory, even though the reference state is a multireference state in terms of the bare $c, c^\dag$ particles. 

This strategy defines a general route to obtain multireference correlation theories 
in the single-reference form. However, a concrete realization
requires the explicit canonical transformation $\hat{U}$ relating $\ket{\Psi_0(c,c^\dag)}$ to a simpler state. 
For generality, we restrict ourselves to $\hat{U}$ which define $\ket{\Psi_0(c,c^\dag)}=\ket{\mathrm{vac}_a}$
as in Eq.~(\ref{eq:ctqpvac}). (This contains 
the particle-number-conserving transformations in
Eq.~(\ref{eq:qp_transformation}) as a special case, because a determinant can always be viewed as a Fermi vacuum
via the particle-hole transformation). Determining $\hat{U}$ 
exactly for a complicated $\ket{\Psi_0(c,c^\dag)}$, such as
a complete active space wavefunction, is of exponential complexity.
Thus, we must introduce approximations. 
We do so by considering approximate canonical transformations corresponding to finite truncations
of the polynomial expansion in Eq.~(\ref{eq:ctqp}).
Then, it is simple to deduce the $\alpha,\beta$ coefficients of $\hat{U}$
from the low-order density matrices of the multireference state.

For example,  consider the lowest-order non-trivial approximation where we truncate Eq.~(\ref{eq:ctqp}) after $\alpha^{(1)}$ and $\beta^{(1)}$,  which corresponds to the Bogoliubov transformation considered in Eq.~(\ref{eq:bogoliubov_general}).
Together with the normalization condition, $\bm{\alpha}^\dag \bm{\alpha} + \bm{\beta}^\dag \bm{\beta}=\bm{1}$, $\alpha^{(1)}$ and $\beta^{(1)}$ are
completely determined by the  single-particle density matrix of  $\ket{\Psi_0}$.
To demonstrate this compactly,
we work with $c, c^\dag$ corresponding to the natural orbital basis of $\ket{\Psi_0}$, i.e.\@  $\langle \Psi_0|c^\dag_p c_q|\Psi_0\rangle = n_p \delta_{pq}$, 
and consider the diagonal spin-restricted Bogoliubov transformation,
\begin{align}
a_p &=  \alpha_{p} c_{p} + s_{p} \beta_{p} c^\dag_{\bar{p}} \ , \notag\\
c_p &= \alpha_{p} a_{p} - s_{p} \beta_{p} a^\dag_{\bar{p}} \ ,
\label{eq:a_bogoliubov_diag}
\end{align}
where we denoted $\alpha \equiv \alpha^{(1)} $ and $\beta \equiv \beta^{(1)}$.
Equating the single-particle density matrices of $\ket{\Psi_0}$ in the $c,c^\dag$ frame and the $a,a^\dag$ frame, we obtain:
\begin{align} 
&\braket{\Psi_0|c^\dagger_{p}c_{p}|\Psi_0} = n_p\nonumber\\
\approx&\braket{\mathrm{vac}_a|\beta_p^2 a_{p}a^\dag_{p}|\mathrm{vac}_a}
=\beta_p^2 \ .
\label{eq:beta}
\end{align}
Thus, $\beta_p = \sqrt{n_p}$, and from the normalization condition $\alpha_p = \sqrt{1 - n_p}$.
Importantly, Eq.~(\ref{eq:beta}) is not an equality, since we truncated the polynomial expansion \eqref{eq:ctqp}, indicating that a complete active space state is {\it not} precisely a Bogoliubov vacuum/BCS state, even
if the two states have the same non-idempotent single-particle density matrix.
The approximation in Eq. \eqref{eq:beta} can be improved by including the
higher orders in Eq.~(\ref{eq:ctqp}). For example, we can determine products such as $\beta^{(1)} \alpha^{(2)}$, etc. from the two-particle density matrix of $\ket{\Psi_0}$. Thus, a complete hierarchy of approximate
canonical transformations corresponding to the full expansion in Eq.~\eqref{eq:ctqp} can be obtained order by order.

The quasiparticle operators $a^{(\dag)}$  acting on the vacuum define
natural excitations to incorporate into the correlation theory.
We first introduce convenient indicial notation. It is usual to choose a convention 
where we divide the operators $a^{(\dag)}$ into 3 classes: (i) fully occupied (core) orbitals with indices $i, j$; (ii) active orbitals with indices $x, y$; and (iii) empty (external) orbitals with indices $a,b$. For the general indices we continue using $p,q,r,s$. 
Correlation theories include excitations of the system between the
core, active and external orbitals. With respect to the quasiparticle vacuum, such
excitations correspond to creating even sets of quasiparticles 
 on top of the quasiparticle vacuum $\ket{\mathrm{vac}_a}\equiv\ket{\mathrm{vac}}$, e.g.
\begin{align}
\ket{ia} = a^\dag_i a^\dag_a \ket{\mathrm{vac}}\ , \ldots \\
\ket{ijxy} = a^\dag_i a^\dag_j a^\dag_x a^\dag_y \ket{\mathrm{vac}}\ , \ldots
\end{align}
Each of the kets generated by these excitations is orthonormal, and the
correlated wavefunction is
\begin{align}
\ket{\Psi} = \ket{\mathrm{vac}} + \sum_{ia} C_{ia} a^\dag_i a^\dag_a \ket{\mathrm{vac}} + \ldots
\end{align}
The expansion coefficients $C$ are formally determined from solving the Schr\"odinger equation in the quasiparticle representation, which requires evaluation of the Hamiltonian matrix elements in the basis of quasiparticles (e.g.\@ $\braket{ia|\hat{H}(a,a^\dag)|\mathrm{vac}}$). As we discussed in the previous section, the quasiparticle Hamiltonian $\hat{H}(a,a^\dag)$ (Eq. \eqref{eq:h_ct}) can be obtained by transforming the original Hamiltonian $\hat{H}(c,c^\dag)$ in Eq. \eqref{eq:h} using the inverse of the polynomial expansion \eqref{eq:ctqp}. Truncating the polynomial expansion \eqref{eq:ctqp} at a low order gives rise to the approximate form of $\hat{H}(a,a^\dag)$, which has a finite (and usually a relatively small) number of terms. As we will show in the next section, in the case of an active-space multireference wavefunction $\ket{\Psi_0}$, the quasiparticle transformation of the Hamiltonian is non-trivial only in the active space, which is usually a relative small part of the orbital space.

\section{Canonically transformed MP2}

We now have all that is necessary to define 
a {\it multireference} dynamic correlation method with
precisely the same form and equations as a {\it single-reference} correlation method.
We refer to these methods as \textit{canonically transformed} correlation methods.
As a simple example, we describe canonically transformed 
M\o ller-Plesset second-order perturbation theory (CT-MP2), where we truncate
the polynomial $a(c,c^\dag)$ at the level of the restricted Bogoliubov 
transformation in the natural orbital basis (Eqs.~(\ref{eq:a_bogoliubov_diag}) and (\ref{eq:beta})). 
Although the Bogliubov transformation is single-particle in form, it captures essential features of the multireference
character of $\ket{\Psi_0}$. Importantly, by construction, it exactly recovers the 
non-idempotent single-particle density matrix of $\ket{\Psi_0}$, and
thus does not require any choice of a ``leading determinant'' in the multideterminant reference wavefunction.
The resulting very simple second-order perturbation
theory from this quasiparticle vacuum thus captures some features of a more traditional
and complicated multireference theory that works with $\ket{\Psi_0}$ directly.

The Bogoliubov Hamiltonian of interest  
$\hat{H}(a,a^\dag)$ contains only up to two-particle (four-index)
terms, which we write explicitly after normal ordering as
\begin{align}
&\hat{H}(a,a^\dag) = E_0 + \sum_{pq} \tilde{t}_{pq} a^\dag_p a_q + \tilde{g}_{pq} a^\dag_p a^\dag_q \notag\\
&+ \frac{1}{4} \sum_{pqrs}  \tilde{v}_{pqrs} a^\dag_p a^\dag_q a_r a_s + \tilde{x}_{pqrs} a^\dag_p a^\dag_q a^\dag_r a_s + \tilde{w}_{pqrs} a^\dag_p a^\dag_q a^\dag_r a^\dag_s \notag\\
&+ h.c.
\label{eq:h_bogoliubov}
\end{align}
Eq. \eqref{eq:h_bogoliubov} can be obtained by inserting the Bogoliubov transformation (\ref{eq:a_bogoliubov_diag}) into the Hamiltonian \eqref{eq:h}. The matrix elements of $\hat{H}(a,a^\dag)$ in Eq.~\eqref{eq:h_bogoliubov} can be determined with at most $\mathcal{O}(M^4)$ cost (where $M$ is the size of the basis set) and are explicitly shown in the Appendix. 

To define the perturbation theory, we choose the zeroth-order Hamiltonian $\hat{H}_0=\tilde{t}_{pq} a_p^\dag a_q$ and semicanonicalize  $\tilde{t}_{pq}\to e_p \delta_{pq}$. 
Then the second-order correlation energy from double excitations is given
by the single-reference formula
\begin{align}
E^{(2)} &= -\frac{1}{4} \sum_{pqrs} 
\frac{\braket{pqrs|\hat{H}|\mathrm{vac}}^2}{e_p+e_q+e_r+e_s} = -\frac{1}{4} \sum_{pqrs}
\frac{\tilde{w}_{pqrs}^2}{D_{pqrs}} \ ,
\label{eq:ctmp2}
\end{align}
where we allow indices $p,q,r,s$ run over $ij\to ab$, $ij\to xy$, and $xy\to ab$ excitations. The spin-orbital expressions for $\tilde{t}_{pq}$ and $\tilde{w}_{pqrs}$ are given in the Appendix.
Eq. \eqref{eq:ctmp2} reduces to that of the standard MP2 in the single-reference limit. 
The CT-MP2 denominator $D_{pqrs} = (e_p+e_q+e_r+e_s)$ can only become zero if the underlying reference state $\ket{\mathrm{vac}}$ is unstable, i.e. if at least one of $e_p < 0$. Contributions
from single excitations also arise from terms such as $\tilde{g}_{pq} a^\dag_p a^\dag_q$ of Eq. \eqref{eq:h_bogoliubov}. 
We do not include such single excitations, as they
vanish in the original theory involving $\ket{\Psi_0}$ (e.g. for an orbital optimized multireference state) but are not strictly
zero here only because $\ket{\mathrm{vac}} \approx \ket{\Psi_0}$. 

It is instructive to analyze the relationship between
CT-MP2 and the standard single-reference MP2 theory.
There are several contributions to the summation in Eq.~(\ref{eq:ctmp2}).
Contributions from core to external orbitals ($ij\to ab$) are equivalent to those in the single-reference case, since $\alpha_i = 0$, $\beta_i = 1$, $\alpha_a = 1$, $\beta_a = 0$ and the denominator $D_{ijab} = (e_a+e_b + e_i + e_j)$ is the standard MP2 denominator, where the sign change from the more usual ($e_a+e_b - e_i - e_j$) is due to the
particle-hole transformation. 
Let us now analyze contributions that arise from active orbitals.
In particular, we consider energy contributions from active to external orbitals ($xy\to ab$), where $x$ and $y$ are weakly occupied ($n_{x,y} \approx 0$), which involve matrix elements $\tilde{w}_{xyab} = v_{xyab}\beta_{x}\beta_{y}$.
For small deviations away from the single-reference Fermi vacuum ($\epsilon$)
we can write $\alpha_x=\cos(\epsilon_x) = 1 - O(\epsilon_x^2)$
 and 
$\beta_x=\sin(\epsilon_x)=\epsilon_x+O(\epsilon_x^3)$. At order $\epsilon^0$, no contributions to the energy arise, since $\beta_x = 0$. The energy contribution at order $\epsilon^1$ is proportional to $(v_{xyab} \epsilon_x \epsilon_y)^2$ scaled by the denominator $(e_a + e_b + e_x + e_y)$. We see that
this is similar to a standard MP2 expression involving the active orbitals
as if they are singly occupied, but with the integral contributions rescaled by a term on the order of the orbital occupancy ($\epsilon_x^2\approx \beta_x^2 = n_x$). Similar analysis can be performed for the strongly occupied active orbitals with $n_{x,y} \approx 1$, where the standard MP2 contributions from $\tilde{w}_{xyab}$ appear at order $\epsilon^0$, while new terms arise 
at order $\epsilon^1$ due to the deviation from single-reference Fermi vacuum.

\begin{figure}[!t]
   \includegraphics[width=0.45\textwidth]{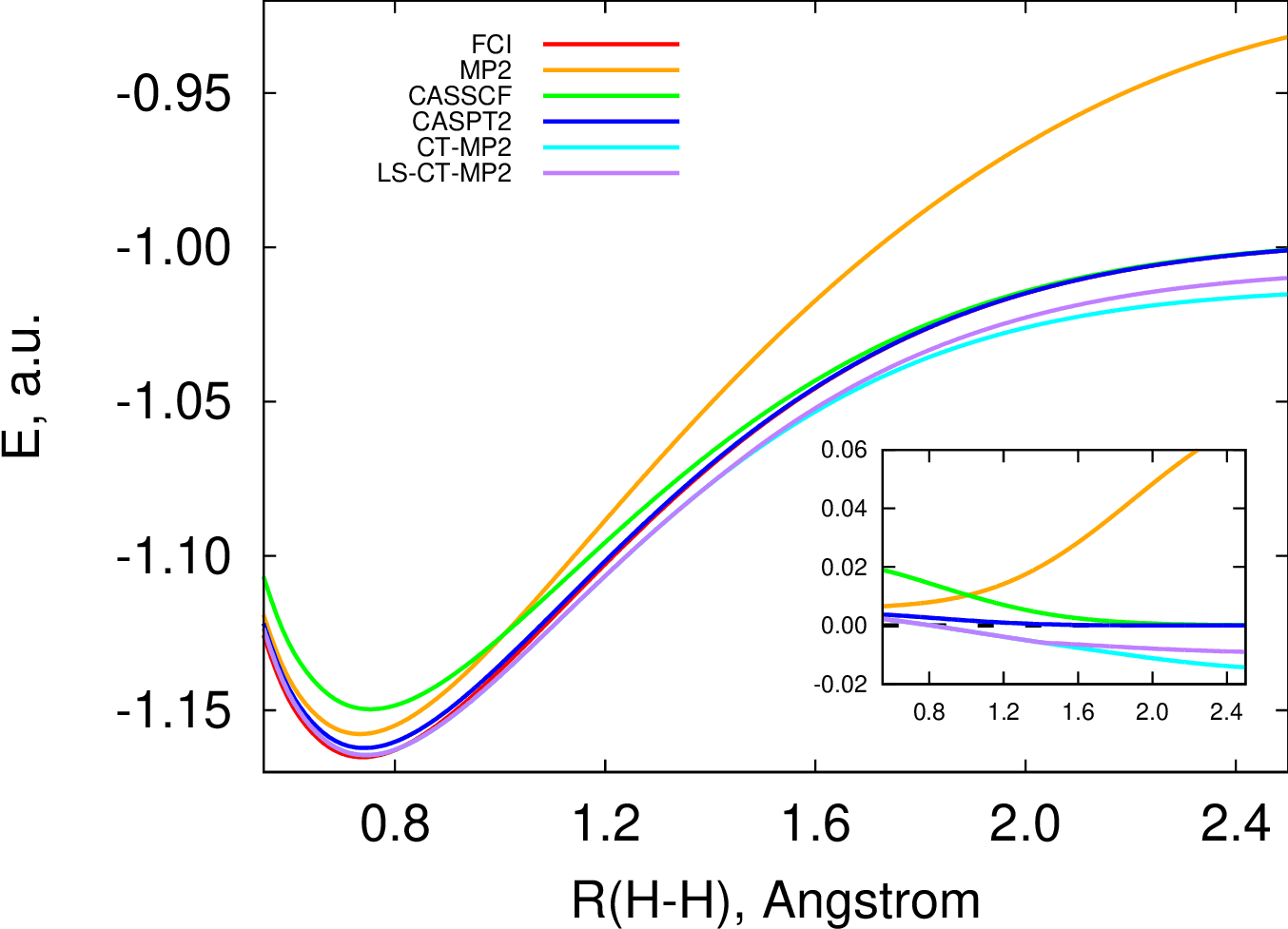}
   \caption{Total energy as a function of the H$_2$ bond length (\mbox{6-31G**} basis set). For CASSCF, CASPT2, and CT-MP2, the (2e, 2o) active space was used. For CT-MP2, results obtained with a level shift are also shown, denoted as LS-CT-MP2 (see text for details). The inset shows deviation of the energy from that of full CI (FCI).}
   \label{fig:H2}
\end{figure}

\begin{figure}[!t]
   \includegraphics[width=0.45\textwidth]{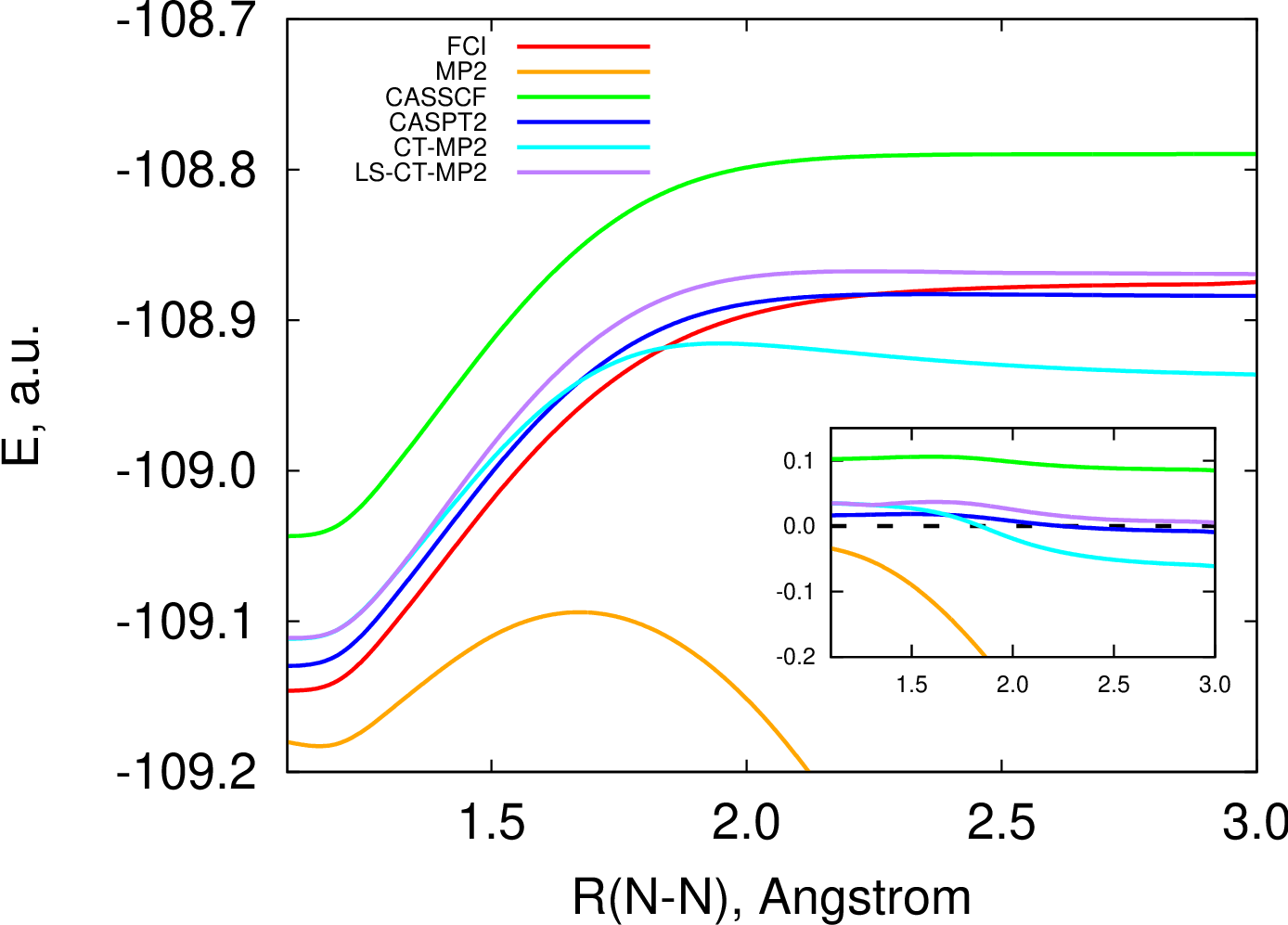}
   \caption{Total energy as a function of the N$_2$ bond length (\mbox{6-311G} basis set). For CASSCF, CASPT2, and CT-MP2, the (6e, 6o) active space was used. For CT-MP2, results obtained with a level shift are also shown, denoted as LS-CT-MP2 (see text for details). The inset shows deviation of the energy from that of full CI (FCI). FCI energies were obtained by freezing $1s$ orbitals of nitrogen. }
   \label{fig:N2}
\end{figure}

\begin{figure}[!t]
   \includegraphics[width=0.45\textwidth]{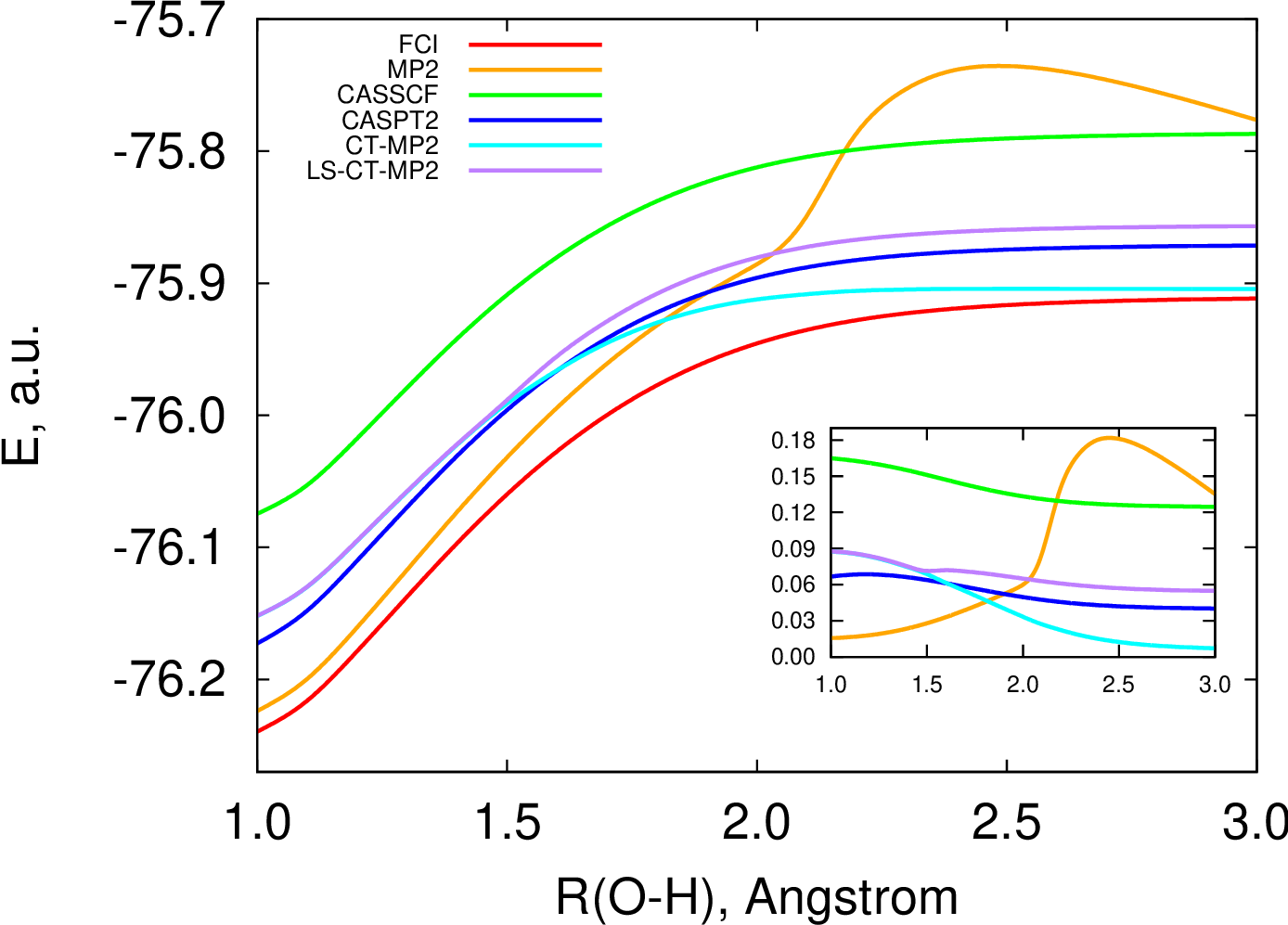}
   \caption{Total energy as a function of the O--H bond length for the symmetric bond dissociation in water molecule (\mbox{cc-pVDZ} basis set). The H--O--H angle was fixed at 109.57$^{\circ}$. For CASSCF, CASPT2, and CT-MP2, the \mbox{(6e, 5o)} active space was used. For CT-MP2, results obtained with a level shift are also shown, denoted as LS-CT-MP2 (see text for details). The inset shows deviation of the energy from that of full CI (FCI). }
   \label{fig:H2O}
\end{figure}

\begin{figure}[!t]
   \includegraphics[width=0.45\textwidth]{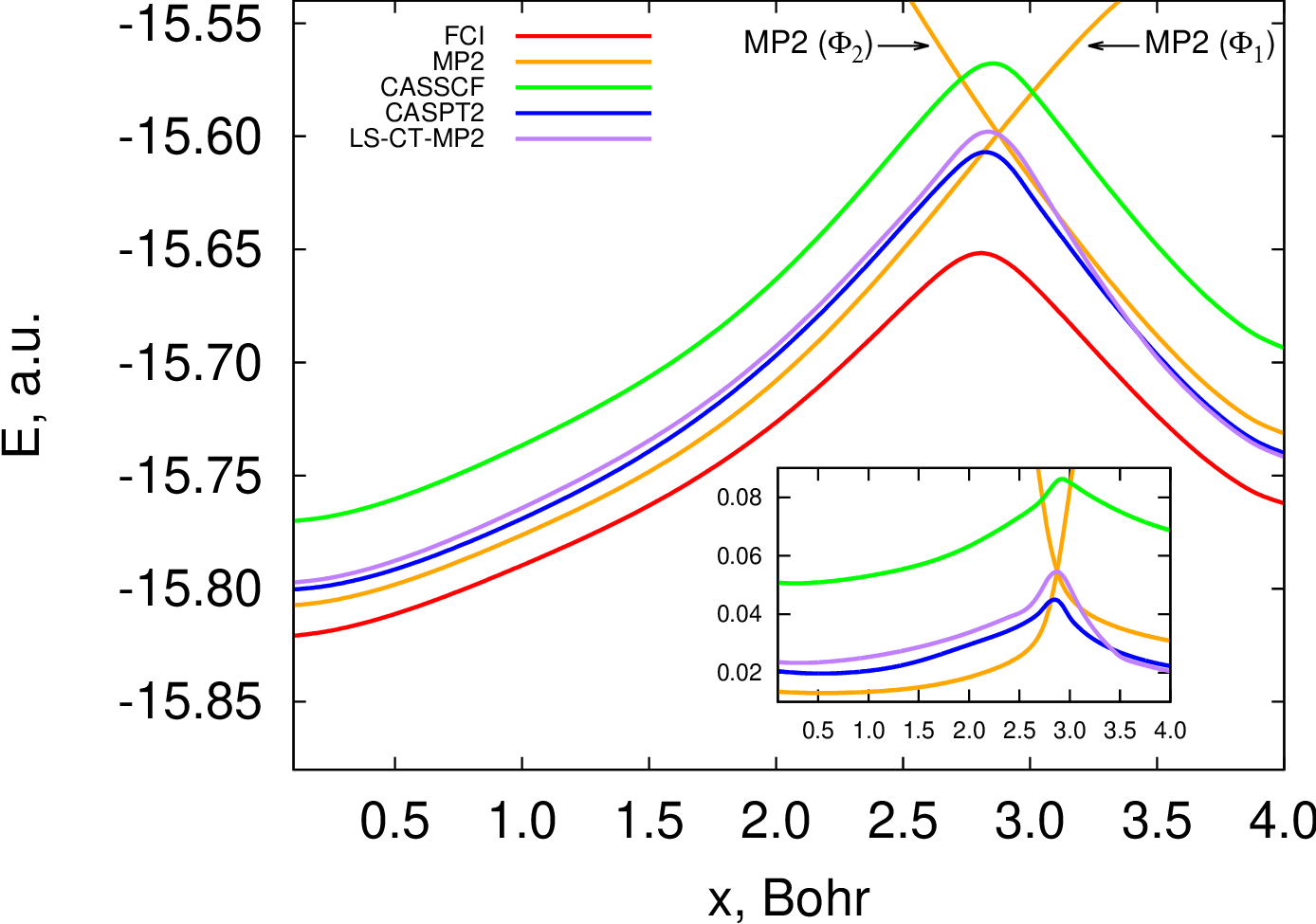}
   \caption{Potential energy curve for the insertion of a beryllium atom into H$_2$ (\mbox{6-311G} basis set).\cite{PurvisIII:1983p835} A beryllium atom is placed at the origin of the coordinate system, while the positions of the hydrogen atoms are defined as $y(x) = \pm(2.54-0.46x)$.\cite{Evangelista:2006p154113} For CASSCF, CASPT2, and CT-MP2, the \mbox{(2e, 2o)} active space was used. For CT-MP2, results were obtained with a level shift, denoted as LS-CT-MP2 (see text for details). The inset shows deviation of the energy from that of full CI (FCI). }
   \label{fig:BeH2}
\end{figure}

\section{Results}
We now demonstrate results of CT-MP2 for the dissociation of H$_2$, N$_2$, symmetric bond-stretching of water, as well as the Be + H$_2$ insertion reaction, relative to full configuration interaction (FCI). We compare the performance of CT-MP2 to that of the conventional MP2 theory and the complete active space second-order perturbation theory (CASPT2). The CT-MP2 total energy was obtained by summing the corresponding CASSCF reference energy and the correlation energy computed via Eq.~\eqref{eq:ctmp2}. For the reference CASPT2 potential energy curves, the energy contributions from single excitations ($i\rightarrow x$, $i\rightarrow a$, and $x\rightarrow a$) were not included.

For the dissociation of H$_2$, N$_2$, and H$_2$O CT-MP2 yields continuous potential energy curves (PECs), whereas single-reference MP2 theory diverges at large bond distances (Figures~\ref{fig:H2} -- \ref{fig:H2O}). Note that the CT-MP2 correlation energy is 
evaluated with the same computational cost as that of MP2 (and is therefore {\it much} less
costly than CASPT2), although it does require the initial CASSCF state. 
CT-MP2 does not perform as well as CASPT2, overestimating the correlation energy at the dissociation limit. This can be traced to the decrease in quality of the Bogoliubov transformed vacuum $\ket{\mathrm{vac}}$ (that lies considerably above the Hartree-Fock determinant in energy) at longer distances, which is indicated by the appearance of the negative active-space eigenvalues $e_x$ of the CT-MP2 zeroth-order Hamiltonian. In the case of N$_2$ dissociation, the poor quality of the quasiparticle vacuum gives rise to an unphysical barrier on the PEC (Figure \ref{fig:N2}). 
The errors near dissociation limit can be attributed to the violation of the particle-number symmetry. In principle,
this can be easily cured by particle number projection, which converts the Bogoliubov vacuum into an antisymmetrized geminal power.\cite{Staroverov:2002p11107,Scuseria:2011p124108,Neuscamman:2012p203001,Neuscamman:2012p045103} 
Practically, the performance of CT-MP2 at long distances can be improved by adding a level shift to the diagonal part of $\hat{H}_0$, such that no negative eigenvalues appear in Eq. \eqref{eq:ctmp2}.\cite{level_shift_note} We define the value of the level shift equal to the largest negative eigenvalue of $\hat{H}_0$. In this case the denominator in Eq. \eqref{eq:ctmp2} is guaranteed to be non-zero and the correlation energy to have a finite value. Figures~\ref{fig:H2} -- \ref{fig:H2O} show the PECs computed using CT-MP2 with a level shift (denoted as LS-CT-MP2). Applying a level shift results in smooth PECs and reduces the CT-MP2 non-parallelity errors from 96 and 81 $mE_h$ to 30 and 33 $mE_h$ for N$_2$ and H$_2$O, respectively (Figures \ref{fig:N2} and \ref{fig:H2O}). The LS-CT-MP2 non-parallelity errors are comparable to those of CASPT2 (25 and 27 $mE_h$ for N$_2$ and H$_2$O).

In the above examples, we did not observe any intruders in the CT-MP2 calculations. We further tested the performance of CT-MP2 for the insertion of a beryllium atom into H$_2$ to form BeH$_2$, a model reaction first studied by Purvis {\it et al.}\cite{PurvisIII:1983p835} Here we employ a modified variant of the original model,\cite{Evangelista:2006p154113} which consists of a beryllium atom placed at the origin of the two-dimensional coordinate system and two hydrogen atoms at positions $y(x) = \pm(2.54-0.46x)$. In the range of $x$ from 0 to 4 $a_0$, the BeH$_2$ wavefunction changes its ground state electron configuration from that of the linear BeH$_2$ ($x< 2.5$ $a_0$, $\ket{\Phi_1}=\ket{(1a_1)^2(2a_1)^2(1b_2)^2}$) to that of the dissociated Be + H$_2$ products ($x> 3$ $a_0$, $\ket{\Phi_2}=\ket{(1a_1)^2(2a_1)^2(3a_1)^2}$). 
In a single-reference treatment, one needs to choose a different dominant
determinant at different bond lengths, thus
one obtains {\it two} distinct single-reference MP2 energy curves (Figure \ref{fig:BeH2}), which cross at $x\approx2.88$ $a_0$.
By contrast, although single-reference in complexity, the Bogoliubov vacuum
exactly reproduces the non-idempotent density matrix of the superpositions
of these two determinants, thus there is only a  {\it single} CT-MP2 curve. 
Nonetheless, discontinuities in the CT-MP2 curve are observed in the region of $2.6 <x< 3.1$ $a_0$. These originate from zero denominators $D_{xyab}$ in Eq. \eqref{eq:ctmp2} due to the appearance of the negative eigenvalues $e_x$. 
In contrast, the level-shifted CT-MP2 (LS-CT-MP2) gives a continuous PEC (Figure \ref{fig:BeH2}), which exhibits a non-parallelity error (34 $mE_h$) comparable to that of CASPT2 (26 $mE_h$). Thus, the very simple LS-CT-MP2 demonstrates the possibility for a simple perturbation theory, with single-reference cost, to provide qualitatively reasonable potential energy curves for complex bond dissociation.

\section{Further connections}

It is appropriate here to explain the connection of our work with the recent work by Rolik and K\'allay in Ref.~\citenum{Rolik:2014p134112}. These authors described a similar strategy to express
multireference theories as single-reference theories in terms of quasiparticles. The main conceptual differences lie in the 
 approximate parametrization and determination of $\hat{U}$. First, Rolik and K\'allay considered only canonical transformations
defined by number-conserving $\hat{U}$. As number-conserving single-particle
canonical transformations are trivial orbital rotations, they had to consider
canonical transformations involving at least two-particle operators, to describe
a non-trivial state (e.g. with a non-idempotent density matrix). However,
the simpler general single-particle (Bogoliubov) transformations we used in 
our calculations above allow any non-idempotent density matrix to be represented,
thus capturing multireference behavior at the single-particle level.
Second (and more importantly), Rolik and K\'allay expressed $\hat{U}$
in exponential form $\hat{U}=\exp \hat{A}$ and
determined the amplitudes of $\hat{A}$  from the full configuration interaction coefficients of the multireference state $\ket{\Psi_0}$. This is a procedure with 
exponential cost. However, as we described above, neither the exponential form nor the full coefficient expansion of $\ket{\Psi_0}$ are necessary to determine the polynomial
expansion of $\hat{U}$ to a finite order. Finally, Rolik and K\'allay described numerical
results for the quasiparticle analogues of coupled cluster theory, while we have focused on perturbation theory.

We further here discuss the connection to the well-known multireference normal ordering
introduced by Mukherjee and Kutzelnigg.\cite{Kutzelnigg:1997p432} They defined the multireference normal ordered
operator pair $\{c_p^\dag c_q\} = c_p^\dag c_q - \gamma_{pq}$, such that
$\langle \Psi_0(c,c^\dag)|\{c_p^\dag c_q\}|\Psi_0(c,c^\dag)\rangle=0$. However, $\{c_p^\dag c_q\}$
is {\it not} a pair of quasiparticle operators. Rather,
\begin{align}
a_p^\dag a_q &= C_0(\alpha,\beta)+C_1(\alpha, \beta) c_p^\dag c_q + C_2(\alpha, \beta) c_pc_q \notag \\
&+ C_3(\alpha, \beta) c_p^\dag c_q^\dag + C_4(\alpha, \beta) c_p^\dag c_q^\dag c_rc_s + \ldots 
\label{eq:mk}
\end{align}
The Mukherjee-Kutzelnigg formalism arises by truncating after the first two terms, 
and setting $C_1=1$, with $C_0$ being fixed by the vacuum expectation value. However,
the general single-particle quasiparticle truncation includes the first four terms. This leads to non-trivial results, as we have seen above.

\section{Conclusions}

In summary, we have formulated a transformation framework to express
multireference theories for dynamic correlation in a simple and natural way, similar to that of the single-reference methods. Our approach works in
a canonically transformed frame of quasiparticles, equating the quasiparticle vacuum to
 the multireference state. The  canonical 
transformation can be practically determined from the low-order density matrices of the multireference
wavefunction. We demonstrated the theory using a low-order
expansion for quasiparticles,  corresponding to a Bogoliubov transformation.
The corresponding canonically transformed second-order M\o ller-Plesset perturbation theory
has single-reference cost (with no high-order density matrices) but is still able
to dissociate multiple bonds, as we demonstrated in the H$_2$, N$_2$,  H$_2$O and BeH$_2$ molecules.
There are many possible extensions of this general framework, and other kinds of multireference dynamic correlation methods can be formulated as canonically transformed versions of the single-reference theories, with exactly the single-reference computational scaling. Further,  higher-level polynomial expansions of the quasiparticle operators 
remain to be explored. 

\vspace{0.5cm}

\section{Appendix: Bogoliubov-Transformed Hamiltonian}
Here we present expressions for the matrix elements of the Bogoliubov-transformed Hamiltonian \eqref{eq:h_bogoliubov} derived using the spin-restricted Bogoliubov transformation in the natural spin-orbital basis (Eq.~\eqref{eq:a_bogoliubov_diag}). For the CT-MP2 method, only the $\tilde{t}_{pq}$ and $\tilde{w}_{pqrs}$ matrix elements are necessary to compute the second-order correlation energy in Eq. \eqref{eq:ctmp2}:
\begin{align}
\label{eq:t_tilde}
\tilde{t}_{pq}  &=
(t_{pq} + \sum_{r} v_{prqr} \beta_{r}^2) \alpha_{p} \alpha_{q}  \notag \\ 
&- 
(t_{\bar{p}\bar{q}} +\sum_{r} v_{\bar{p}r\bar{q}r}  \beta_{r}^2 ) \beta_{\bar{p}} \beta_{\bar{q}} s_{\bar{p}} s_{\bar{q}} 
\notag \\
&+\frac{1}{2}\sum_{r} (v_{p\bar{q}\bar{r}r} \alpha_{p} \beta_{\bar{q}} s_{\bar{q}} 
+v_{q\bar{p}\bar{r}r} \alpha_{q} \beta_{\bar{p}} s_{\bar{p}} )\alpha_{r} \beta_{\bar{r}} s_{\bar{r}} \ ,
\end{align}
\begin{align}
\label{eq:w_tilde}
\tilde{w}_{pqrs}  &=
v_{pq\bar{s}\bar{r}} \alpha_{p} \alpha_{q} \beta_{\bar{r}} \beta_{\bar{s}} s_{\bar{r}} s_{\bar{s}} \ .
\end{align}
Note that the matrix elements in Eqs. \eqref{eq:t_tilde} and \eqref{eq:w_tilde} can be computed with at most $\mathcal{O}(M^4)$ scaling where $M$ is the size of the basis set. 
Expressions for other matrix elements are shown below:
\begin{align}
\label{eq:e_0}
E_{0}  
&=\sum_{p} t_{pp} \beta_p^2
+ \frac{1}{2} \sum_{pq} v_{pqpq}\beta_p^2\beta_q^2 \notag \\
&+ \frac{1}{4} \sum_{pq} v_{p\bar{p}q\bar{q}}\beta_p\alpha_{\bar{p}}\beta_q\alpha_{\bar{q}}s_p s_q \ ,
\end{align}
\begin{align}
\label{eq:g_tilde}
\tilde{g}_{pq}  
&= (t_{p\b{q}} + \sum_{r} v_{pr\b{q}r} \beta_{r}^2) \alpha_{p} \beta_{\b{q}} s_{\b{q}}  \notag \\
&+ \frac{1}{4} \sum_{r} (v_{pqr\b{r}}\alpha_p\alpha_q + v_{\b{q}\b{p}r\b{r}}\beta_{\b{p}}\beta_{\b{q}}s_{\b{p}}s_{\b{q}}) \alpha_{\b{r}} \beta_{r} s_r  \ ,
\end{align}
\begin{align}
\label{eq:v_tilde}
\tilde{v}_{pqrs}  
&= v_{pqsr} \alpha_{p} \alpha_{q} \alpha_{r} \alpha_{s}  \notag \\
&+ v_{\b{p}\b{q}\b{s}\b{r}} \beta_{\b{p}} \beta_{\b{q}} \beta_{\b{r}} \beta_{\b{s}} s_{\b{p}} s_{\b{q}} s_{\b{r}} s_{\b{s}} \notag \\
&+ 4 v_{p\b{r}\b{q}s} \alpha_{p} \beta_{\b{r}} \beta_{\b{q}} \alpha_{s} s_{\b{r}} s_{\b{q}}  \ ,
\end{align}
\begin{align}
\label{eq:x_tilde}
\tilde{x}_{pqrs}  
&= 2 v_{pqs\b{r}} \alpha_{p} \alpha_{q} \alpha_{s} \beta_{\b{r}} s_{\b{r}}  \notag \\
&+ 2 v_{p\b{s}\b{r}\b{q}} \alpha_{p} \beta_{\b{s}} \beta_{\b{r}} \beta_{\b{q}} s_{\b{s}} s_{\b{r}} s_{\b{q}} \ .
\end{align}
The remaining terms in Eq. \eqref{eq:h_bogoliubov} are Hermitian conjugates of Eqs. \eqref{eq:w_tilde}, \eqref{eq:g_tilde}, and \eqref{eq:x_tilde}. In the single-reference limit, Eqs. \eqref{eq:t_tilde} -- \eqref{eq:x_tilde} reduce to matrix elements of the standard single-reference normal-ordered Hamiltonian. 

\section{Acknowledgements}
This work was supported by the US Department of Energy, Office of Science through
Award DE-SC0008624, and Award DE-SC0010530. The authors would like to thank Bo-Xiao Zheng and Francesco Evangelista for helpful discussions.


\begin{thebibliography}{10}
\bibitem{level_shift_note}
We note that an optimized level shift is contained by default within the
  definition of the standard CASPT2 method.
\end{thebibliography}
\end{document}